\documentclass[aip,reprint]{revtex4-1}

\usepackage{amsthm}
\usepackage{amssymb}
\draft 
\usepackage{color}
\usepackage[colorlinks, citecolor=blue]{hyperref}
\usepackage[figuresright]{rotating}
\usepackage{graphicx}
\begin{document}


\title{Effect of electrode geometry on photovoltaic performance of polymer solar cells} 



\author{Meng Li}
\affiliation{Department of Physics, Henan Normal University, Xinxiang 453007, China}
\affiliation{Henan Key Laboratory of Photovoltaic Materials, Xinxiang, 453007, China}
\author{Heng Ma}
\altaffiliation{Electronic mail:hengma@henannu.edu.cn.}
\affiliation{Department of Physics, Henan Normal University, Xinxiang 453007, China}
\affiliation{Henan Key Laboratory of Photovoltaic Materials, Xinxiang, 453007, China}
\author{Zhao-kui Wang}
\affiliation{Institute of Functional Nano \& Soft Materials (FUNSOM), Soochow University, Suzhou, Jiangsu, 215123, China}
\author{Chuan-kun Wang}
\affiliation{Department of Physics, Xingyi Normal University of Nationalities, Xingyi, 562400, China}
\author{Yu-rong Jiang}
\affiliation{Department of Physics, Henan Normal University, Xinxiang 453007, China}
\affiliation{Henan Key Laboratory of Photovoltaic Materials, Xinxiang, 453007, China}
\author{Ning Liu}
\affiliation{Department of Physics, Henan Normal University, Xinxiang 453007, China}
\affiliation{Henan Key Laboratory of Photovoltaic Materials, Xinxiang, 453007, China}




\begin{abstract}
This paper investigates the impact of electrode geometry on the performance of polymer solar cells (PSCs). Four types of negative electrodes with equal area (0.09 cm$^{2}$) but different shape (round, oval, square, and triangular) are evaluated with respect to short-circuit current density, open-circuit voltage, fill factor, and power conversion efficiency of PSCs. The results show that the device with round electrodes gives the best photovoltaic performance; in contrast, the device with triangular electrodes reveals the worst properties. Maximum almost twice increase in PCE with round electrode is obtained in the devices compared with that of the triangular electrode. As a conclusion, the electrode boundary curvature has a strong influence on the performance of PSCs. The larger curvature, i.e. the sharper electrodes edge, maybe is a negative effector on exciton separation and carrier transport in PSC system.
\end{abstract}

\pacs{}

\maketitle 

\section*{1. Introduction}  

Development of new energy sources is gaining importance and has become an issue of common concern. Because solar energy is environment-friendly energy, its utilization is being investigated intensively by many researchers and is supported by many governments. The effective use of solar energy has become a very active research field, especially in the field of solar cells. Polymer solar cell (PSC) represent one of the most popular research topics because of their numerous advantages such as lightweight, mechanical flexibility, ecofriendliness, and low-cost production of electronic devices. The performance of PSCs significantly improved in the 1990s itself because of the introduction of bulk heterojunction structures. Recent advances indicate that the power conversion efficiency (PCE) of PSCs has been greatly advanced to reach a level of 9\textsf{\%} \cite{he2012enhanced,dou2012tandem,Li2013Research}.

　
　　　The most representative research on bulk heterojunction PSCs is based on poly(3-hexylthiophene) (P3HT) and [6,6]-phenyl-C$_{61}$-butyric acid methyl ester (PCBM) with the corresponding PCE up to 5\textsf{\%}. Many researchers have been using different production processes and device structures to improve PCE, such as spin coating, dripping, inkjet printing, and vacuum deposition. So far, some effective achievements have been obtained. These developments are reflected in the active layer, buffer layer, electrode materials, production process, and changes in the cell structure \cite{guo2010thieno,bauer2012zno,liang2012effects,schmidt2012structure,Ma2011degradation}, among others. In addition to the conventional structure of solar cells, researchers have also introduced inverted and tandem structures to further improve the PCE with promising results \cite{you2012metal,larsen2012roll}.

　　　Electrode materials and processes are the main factors in cell fabrication. In recent years, research in these areas has become of widespread concern \cite{li2012polymer,sista2010high,cai2010polymer}. However, there are only a few studies on the influence of the electrode geometry on the cell performance \cite{lee2012polymer,Kuwabara20121136}. A series of works on optimizing the electrode geometry of dye-sensitized solar cells have been reported \cite{lee2009effect,ito2006photovoltaic,park2007v}. In contrast, for PSCs, to the best of our knowledge, only a few reports about the impact of the electrode geometry are found. Generally, rectangle is the common shape of PSCs electrodes. Separated excitons, i.e., the early generated charge carriers will gather on the electrode to build an electric field. For the next carriers, exciton generation and separation, as well as carrier distribution and transmission will be affected by the built-in electric field. Therefore, the electrode geometry and boundary that determine the built-in electric field are the key factors for determining the distribution of excitons and carriers that influence the performance of PSCs.

　　　In this work, four types of PSCs- round, oval, square and triangle patterned -are fabricated by the evaporation of negative electrodes with same area (0.09 cm$^{2}$) and are evaluated. By analyzing the physical mechanism by which the electrode geometry affects the electric-field distribution in the cells, a novel influence factor of electrode on the performance of PSCs is discussed. This work might provide some useful reference for determining the efficiency of PSCs.

\section*{2. Experiment}
\label{2}

\subsection*{2.1 Experimental materials}
\label{5}
　　Regioregular active materials P3HT and PCBM were purchased from Luminescence Technology Corp. and Solenne BV. As an aqueous dispersion Poly(3,4-ethy-lenedioxythiophene): poly(4-styrenesulfonate) (PEDOT: PSS) was purchased from Bayer AG (CLEVIOS$^{\textrm{TM}}$ P VP AI 4083). All the chemicals were used as received without further purification. The chemical structures of the materials are shown in \textcolor[rgb]{0.00,0.00,1.00}{Fig. 1}. The indium tin oxide (ITO) substrate with a sheet resistance of average 15 $\Omega/\Box$ and transmission 85\textsf{\%} was purchased from Kaivo.

\begin{figure}[htbp]
\centering
\includegraphics[totalheight=1in]{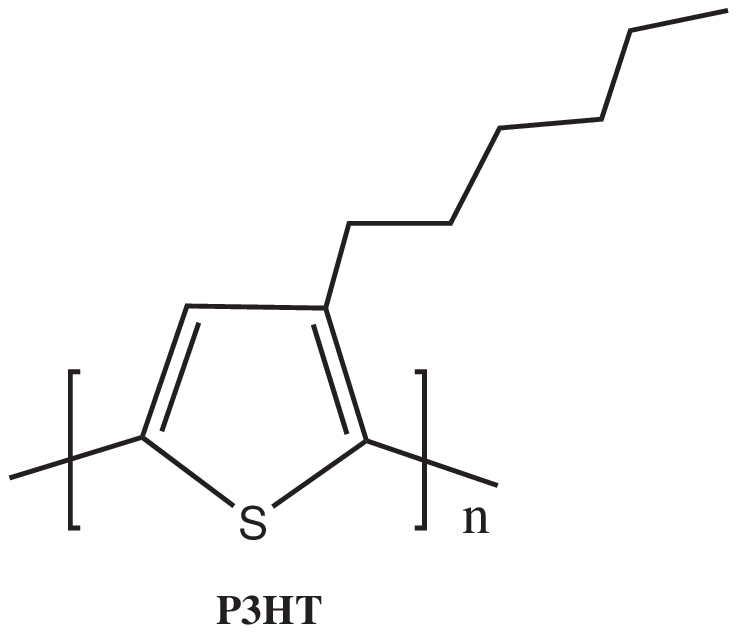}
\hspace{1in}
\includegraphics[totalheight=1in]{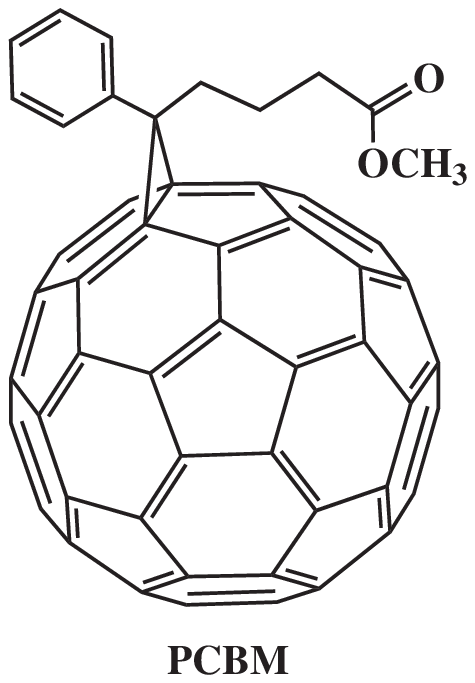}

\includegraphics[totalheight=1in]{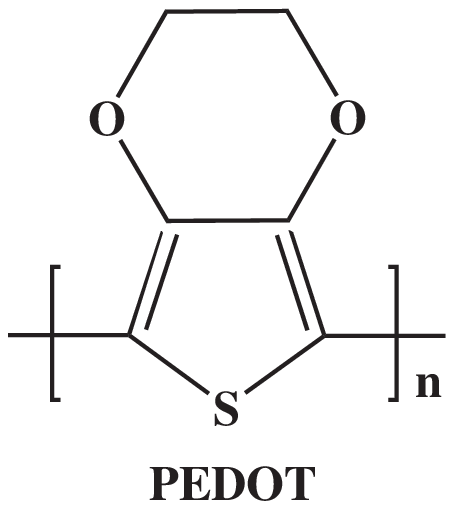}
\hspace{1in}
\includegraphics[totalheight=1in]{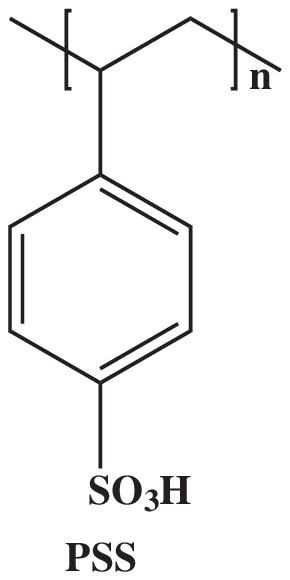}
\caption{The chemical structures of P3HT, PCBM, PEDOT and PSS.}
\label{F1}       
\end{figure}
\subsection*{2.2 Device fabrication and characterization}
\label{6}
　　In the experiment, ITO substrate was cut into 1.8cm $\times$ 1.8cm chips. Four kinds of cells were produced on one substrate so as to ensure the devices of different geometry having same preparation conditions. In order to eliminate the disturbance caused from uncertainties, the simplest cell structure, GLASS/ITO/PEDOT:PSS/ P3HT: PCBM/Al, is employed in the experiment.

　　 Because the etching process of ITO anode is easy to result in concave and convex defect of the electrode edge, the active and buffer layer are spin coated straightly on ITO substrate, firstly. Then, aluminum cathode is evaporated on the active layer using a mask with equal area but different shape (round, oval, square, and triangular). In addition to the aluminum electrode was  evaporated in vacuum, the other production processes, including current voltage (J-V) measurement, were manufactured in atmosphere environment.

　　The specific processes and condition parameters adopted in this work are listed as follow.

　　Substrates cleaning: aqueous detergent / deionized water / acetone / alcohol / deionized water (10 min each process, at 30 $^{\circ}\mathrm{C}$, in ultrasonic basin). The cleaned substrates were dried in air.

　　PEDOT: PSS spin coating (thickness: $\sim$50 nm): 1500 rpm 15 s / 3000 rpm 45 s / 110 $^{\circ}\mathrm{C}$ annealed 10 min (in air).

　　Preparation of active layer solution: P3HT 10 mg, PCBM 9 mg, 1, 2-dichlorobenzene 1 mL are mixed and stirred at 60 $^{\circ}\mathrm{C}$ for 24 h.

　　Spin coating of active layer (thickness: $\sim$180 nm): 500 rpm 15 s / 700 rpm 30 s / 110 $^{\circ}\mathrm{C}$ annealed 10 min (in air).

　　Al electrode evaporation (thickness: $\sim$100 nm): 0.09 cm$^{2}$ pattern mask (4 types) / 99.995\textsf{\%} Aluminum / vacuum 5 $\times$ 10$^{-4}$ Pa / 30 min.

　　Solar simulator: 1000 W Xenon lamp, AM 1.5 G, 100 mW/cm$^{2}$ (Abet Technologies Cop.).

　　J-V measurement: four wire method, Keithley 2400 source meter.

 In the measuring process of J-V curve, another mask which is curved according to the four electrode pattern was attached on the front of the cell. It can insure the illumination to shine on the effect region of the cell, i.e. the evaporated aluminum electrode region. The anode lead was linked from the edge of ITO substrate; the cathode leaded from the center of the Al island electrode using a section of copper wire and silver sol.

\section*{3. Results and Discussion}
\label{3}

\subsection*{3.1 Geometric distribution of electric field between electrodes}
\label{7}
　　Because the purpose of our present work is to study the influence of the electrode shape on the cell performance, it is necessary to first discuss the geometric distribution of electric field between the electrodes of PSCs. According to the charge distribution in bulk heterojunctions, we can qualitatively analyze the potential and electric-field distribution.

　　It is generally known that the charge distribution of a charged body correlates with the curvature of the pattern. In bulk heterojunction PSCs, early generated carriers reach the electrode to form an initial built-in electric field that will influence the subsequent exciton separation and carrier transport. A larger curvature, i.e. sharper electrode edges, can gather more carriers to form a strong built-in electric field. Therefore, the electric-field distribution of the cells will be nonuniform and asymmetrical because of the different geometric shapes (round, oval, square, and triangular). \textcolor[rgb]{0.00,0.00,1.00}{Fig. 2} depicts the schematics of the electric-field distributions corresponding to the different electrode patterns used in this study.

\begin{figure}[htbp]
\centering
\includegraphics[totalheight=1in]{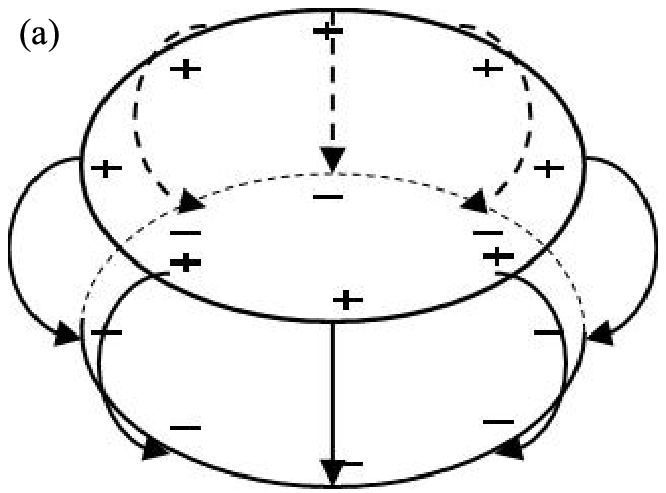}
\hspace{0.4in}
\includegraphics[totalheight=1in]{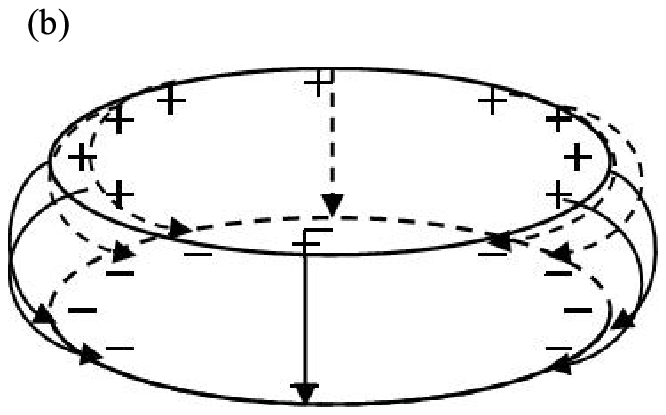}

\includegraphics[totalheight=1in]{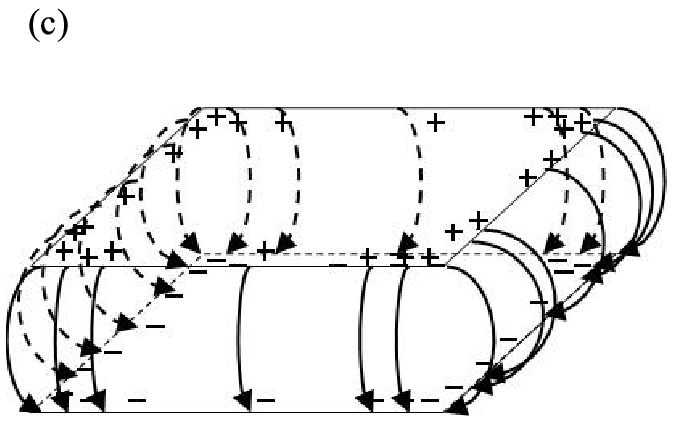}
\hspace{0.4in}
\includegraphics[totalheight=1in]{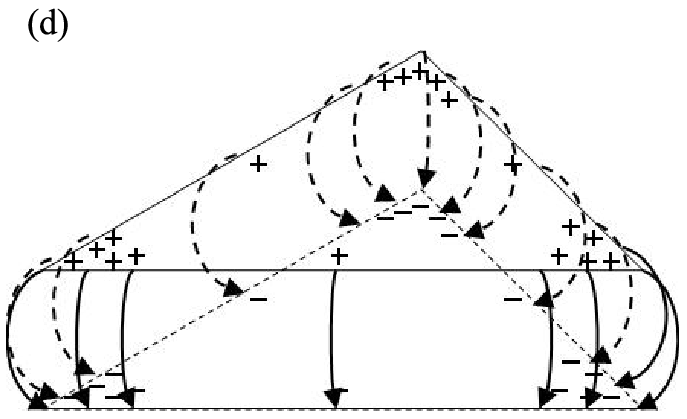}
\caption{Schematics of the electric-field distributions with different electrode geometry (a) round electrode (b) oval electrode (c) square electrode (d) triangular electrode.}
\label{F2}       
\end{figure}
\subsection*{3.2 Influence of electrode geometry on cell performance}
\label{8}
　　In order to more accurately explore the influence of the electrode shape, a shadow mask corresponding to the shape of each cathode was firmly attached on the front of PSCs to ensure that the anode and cathode receive the same illumination during the measurements. \textcolor[rgb]{0.00,0.00,1.00}{Fig. 3} shows the J-V characteristics of the devices with different electrode geometries under white light illumination. From \textcolor[rgb]{0.00,0.00,1.00}{Fig. 3}, the short-circuit current density (J$_{\textrm{sc}}$) of the cell equipped with a round electrode (R cell) is the largest one and is approximately two times greater than that of the triangular electrode (T cell). Meanwhile, the cell with an oval electrode (O cell) also possesses larger current density compared with the T cell and square cell (S cell) whose electrode edges are angular. Therefore, one can conclude that the curvature of the electrode is one of the determinants of J$_{\textrm{sc}}$ in PSCs. On the other hand, regarding the open-circuit voltage (V$_{\textrm{oc}}$), few differences, which are caused by the curvature of the electrode, can be found among the four types of PSCs.
\begin{figure}[htbp]

\centering
\includegraphics[width=0.45\textwidth]{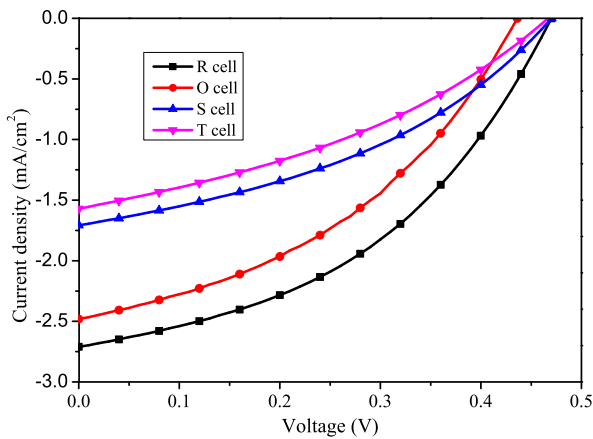}
\caption{Current density-voltage (J-V) characteristics of
　　PSCs with different electrode geometry. }
\label{F3}       
\end{figure}

　　J$_{\textrm{sc}}$ and V$_{\textrm{oc}}$ of the cells can be obtained from the J-V curves of our measurement. The fill factor (FF) and PCE of the devices are calculated based on \textcolor[rgb]{0.00,0.00,1.00}{Eq. 1} and \textcolor[rgb]{0.00,0.00,1.00}{Eq. 2}.

　　According to the general theory of organic semiconductor, the FF of the cell is defined as the maximum power, and it is expressed as
\begin{eqnarray}
 \textrm{FF}=\frac {\textrm{P}_{\textrm{max}}}{\textrm{I}_{\textrm{sc}}\textrm{V}_{\textrm{oc}}}
 =\frac{\textrm{I}_{\textrm{max}}\textrm{V}_{\textrm{max}}}{\textrm{I}_{\textrm{sc}}\textrm{V}_{\textrm{oc}}},
\end{eqnarray}

Where V$_{\textrm{max}}$ and I$_{\textrm{max}}$ are voltage and current at maximum power point, respectively.

  PCE of PSCs is defined as the ratio of device's maximum output power (P$_{\textrm{max}}$) and the irradiation power (P$_{\textrm{in}}$=100 mW/cm$^{2}$). That is,
 \begin{eqnarray}
 \eta=\frac {\textrm{P}_{\textrm{max}}}{\textrm{P}_{\textrm{in}}}
 =\frac{\textrm{FF}\times \textrm{V}_{\textrm{oc}}\times \textrm{I}_{\textrm{sc}}}{\textrm{P}_{\textrm{in}}},
\end{eqnarray}

　　The resulting parameters are summarized in \textcolor[rgb]{0.00,0.00,1.00}{Table I}. It is obvious that FF and PCE are strongly influenced by the electrode geometry, similar to the short-circuit current.

　　
 For comparison, the relationships between J$_{\textrm{sc}}$, V$_{\textrm{oc}}$, FF, PCE, and the electrode geometry are summarized in \textcolor[rgb]{0.00,0.00,1.00}{Fig. 4}. The experimental results show that the geometry of the electrodes has an apparent impact on the cell performance.

\begin{figure*}[htbp]

\centering

\includegraphics[totalheight=2in]{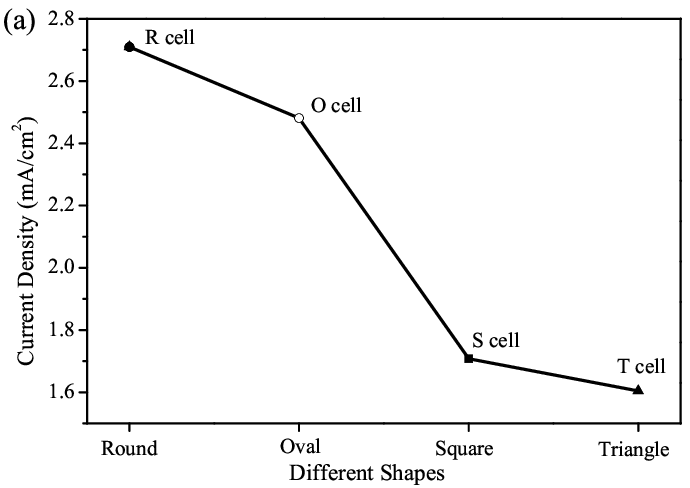}
\hspace{0.1in}
\includegraphics[totalheight=2in]{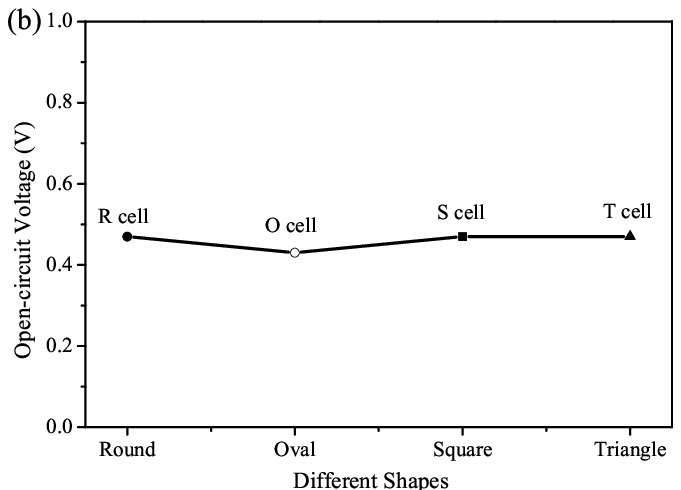}

\includegraphics[totalheight=2in]{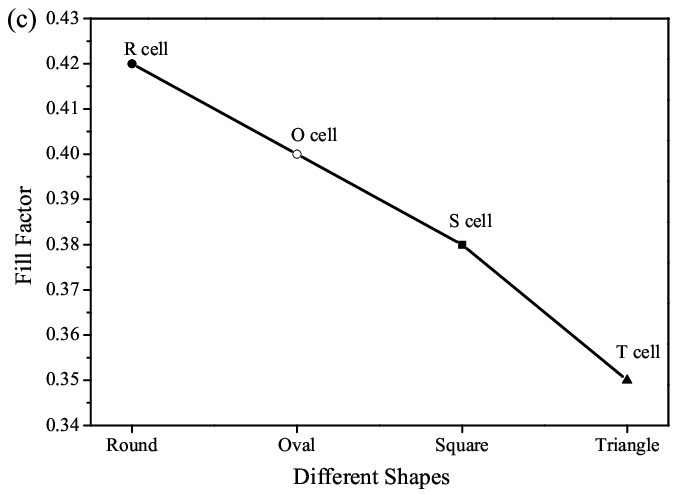}
\hspace{0.1in}
\includegraphics[totalheight=2in]{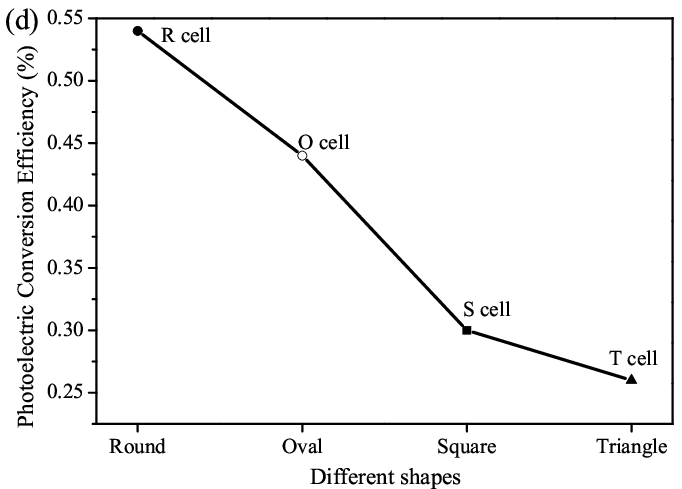}
\caption{The relationship between the different electrode geometry of PSCs and device performance (a) Current density, (b) Open-circuit voltage, (c) fill factor and (d) Photoelectric conversion efficiency.}
\label{F4}       
\end{figure*}
\textcolor[rgb]{0.00,0.00,1.00}{Fig. 4}(a) depicts Jsc as a function of the electrode geometry, which follows the following order:

　　          \qquad  \qquad           R cell $>$ O cell $>$ S cell $>$ T cell.

　　The values of J$_{\textrm{sc}}$ decrease with increasing sharpness of the electrode contour. As discussed in \textcolor[rgb]{0.00,0.00,1.00}{Fig. 3}, J$_{\textrm{sc}}$ will be smaller as the curvature of the electrode becomes larger, i.e., the angles of the electrode become sharper.

Generally, there are numerous complex influencing factors that determine the J$_{\textrm{sc}}$ of PSCs, whose mechanism, however, has not been completely understood. J$_{\textrm{sc}}$ is mainly determined by the quantum efficiency of the light absorption process, collection efficiency of the carriers at the electrode, and internal electrical resistance \cite{nikitenko2007non, monestier2007modeling}. Moreover, the thickness of the active layer, carrier mobility of the active layer-materials, especially, the interfacial properties between the electrode and the active material can also affect J$_{\textrm{sc}}$.

\begin{table}[htbp]
\caption{Photovoltaic performance characteristics of PSCs with different electrode geometry.}
\label{tab:I}       
\centering
\begin{tabular}{ccccccc}
\noalign{\smallskip}\hline
\hline\noalign{\smallskip}
\textrm{Device} & $\emph{\textrm{V}}_{\textrm{oc}}$ & $\emph{\textrm{J}}_{\textrm{sc}}$& \textrm{FF} & $\textrm{PCE}$
& $\emph{\textrm{R}}_{\textrm{s}}$ & $\emph{\textrm{R}}_{\textrm{p}}$\\

  &$(\textrm{\textrm{V}})$&$(\textrm{mA}/\textrm{cm}^{2})$& &$(\textsf{\%})$&$(\textrm{$\Omega$})$&$(\textrm{$\Omega$})$\\
\noalign{\smallskip}\hline\noalign{\smallskip}
R cell&0.47&2.70&0.42&0.54&882.0&7001.4\\
O cell&0.43&2.48&0.40&0.44&976.2&6013.9\\
S cell&0.47&1.70&0.38&0.30&1624.4&7594.0\\
T cell&0.47&1.57&0.35&0.26&1959.7&6608.1\\
\noalign{\smallskip}\hline
\hline\noalign{\smallskip}
\end{tabular}
\end{table}
　　In fact, the electrode geometry, where an uneven potential gradient can form an inhomogeneous electric field, might be another important factor that is reflected in the carrier mobility and carrier-collection efficiency. As mentioned above, the heterojunction causes a random carrier distribution within the active layer during the manufacturing process of a bulk heterojunction PSC. The early generated carriers reach the electrode to form an initial built-in electric field that will influence the exciton separation and carrier transport. A larger curvature of the electrode, such as square and triangular pattern, can lead to a larger accumulation of carriers that, in turn, forms a strong built-in electric field in the angular area. Therefore, a large force will be exerted on newly generated excitons and carriers will gather around the strong built-in electric field. The excitons might be spatially confined and their separation rate will decrease because of the nonuniform electric field force. Meanwhile, this electric field force will lead to a transport imbalance between the holes and electrons and even increase the probability of the carrier recombination. On the other hand, the inside of the active layer is prone to the accumulation of space charge, forming a space-charge layer that also easily leads to a carrier recombination while reducing the charge-collection efficiency.

　As a result, the geometric shape of the electrodes can strongly affect the exciton separation, carrier transport, mobility of the carriers, and carrier-collection efficiency. Hence, J$_{\textrm{sc}}$ is lower in case of S and T cells. This is the main physical mechanism by which the electrode geometry influences the performance of PSCs.

　　\textcolor[rgb]{0.00,0.00,1.00}{Fig. 4}(b) shows the relationship between V$_{\textrm{oc}}$ and electrode geometry. Different cells acquire almost equal values of V$_{\textrm{oc}}$, except in case of the O cell. This indicates only a slight influence of the curvature of the electrode on V$_{\textrm{oc}}$. V$_{\textrm{oc}}$ mainly depends on the highest occupied molecular orbital (HOMO) of the donor material and the lowest unoccupied molecular orbital (LUMO) of the receptor material \cite{rand2007offset}. Therefore, its correlation with the electrode geometry is small.

　　\textcolor[rgb]{0.00,0.00,1.00}{Fig. 4}(c) illustrates the relationship between FF and electrode geometry. These two factors exhibit correlation similar to J$_{\textrm{sc}}$. The FF of R cell is the largest, whereas that of S cell is the smallest. With increase in the sharpness of the electrode edges, the corresponding FF shows a decreasing trend. In the case of PSCs, FF is mainly affected by the transport properties of holes and electrons \cite{kroon2008small}. As already mentioned, the different electric-field distributions caused by the different electrode geometries determines the transport and mobility of the carriers.

　　\textcolor[rgb]{0.00,0.00,1.00}{Fig. 4}(d) summarizes the measurement results of the PCEs for the cells with round, oval, square, and triangular electrode shapes. Because the simplest cell structure is employed, and electrode fabrication and J-V measurements are performed under the atmospheric environment, the absolute PCE values are lower than those reported in some references. Meanwhile, Al cathode conduction system increases the potential of a large series resistance in the measuring process. Table 1 lists the data of the series resistor $\emph{\textrm{R}}_{\textrm{s}}$ and parallel resistance $\emph{\textrm{R}}_{\textrm{p}}$, respectively. This is also an important reason why the PCE of our cells is so lower. However, the corresponding values of our cells are credible and the compared results are convincing. PCE of PSCs is defined by the ratio of the device's maximum output power to the incident irradiation power. \textcolor[rgb]{0.00,0.00,1.00}{Eq. 2} shows that PCE is proportional to V$_{\textrm{oc}}$, J$_{\textrm{sc}}$, and FF. As discussed previously, though the electrode geometry does not affect V$_{\textrm{oc}}$, it significantly affects J$_{\textrm{sc}}$ and FF. Therefore, R and O cells possess PCE values larger than T and S cells that have angular electrode profiles and thus display smaller PCE. As a result, the electrode boundary curvature reveals a negative correlation with PSC performance.

\section*{4. Conclusion}
\label{4}
In conclusion, the influence factor of the electrode geometry on the photovoltaic performance of PSCs is discussed. Under the identical electrode areas and manufacturing conditions, the electrode geometry shows a significant impact on J$_{\textrm{sc}}$, FF, and PCE; however, it hardly affects V$_{\textrm{oc}}$. By analyzing the formation process of the electric field in the electrode, we infer that angular electrodes form uneven electric fields that block the carrier transport and reduce the carrier mobility, even in case of excitons. We expect that this work on the electrode geometry is useful for evaluating the performance of PSCs not only in laboratory studies but also in commercial applications.

\section*{Acknowledgement}
This work is supported by the National Natural Science Foundation of China under grant No. 11074066.

%



\end{document}